\documentclass[12pt]{article}
\usepackage[dvips]{graphicx}
\usepackage{mathrsfs}
\usepackage{amsthm} 
\usepackage{amsmath}
\usepackage{amssymb}
\usepackage{amsfonts}
\usepackage{float}
\usepackage{yhmath}
\usepackage{wrapfig}
\usepackage[dvips]{color}
\usepackage[dvips]{colortbl}
\usepackage{cite}
\usepackage{array}
\usepackage{tabularx}
\usepackage[sectionbib]{chapterbib}
\usepackage{threeparttable}
\usepackage{chemarrow}
\usepackage{framed}
\usepackage{ascmac}
\definecolor{shadecolor}{gray}{0.90}
\usepackage[small, bf, labelsep=quad]{caption}
\usepackage{cases}

\newcommand{\CF}{C_{\textit{\textsf{\hspace{-0.3mm}F}}}}

\newcommand{\HG}{\hspace{1.3mm}\Hat{\textit{\textsf{\hspace{-1.3mm}G}}}\hspace{0.5mm}}
\newcommand{\I}{I}
\newcommand{\II}{I\hspace{-0.5mm}I}

\definecolor{mycolor1}{rgb}{1,1,0.7}
\definecolor{mycolor2}{rgb}{0.9,1,1}
\definecolor{mycolor3}{cmyk}{0,0,0,0.113}
\definecolor{mycolor4}{cmyk}{0.086,0,0,0}

\begin{document}

\begin{center}
{\huge The Equation of Excluded Volume Effects }
\end{center}
\begin{center}
{\Large A Minor Improvement in the Evaluation of the Inhomogeneity Term}
\end{center}

\vspace*{5mm}
\begin{center}
\large{Kazumi Suematsu\footnote{\, The author takes full responsibility for this article.}, Haruo Ogura$^{2}$, Seiichi Inayama$^{3}$, and Toshihiko Okamoto$^{4}$} \vspace*{2mm}\\
\normalsize{\setlength{\baselineskip}{12pt} 
$^{1}$ Institute of Mathematical Science\\[1mm]
Ohkadai 2-31-9, Yokkaichi, Mie 512-1216, JAPAN\\
E-Mail: suematsu@m3.cty-net.ne.jp, ksuematsu@icloud.com  Tel/Fax: +81 (0) 593 26 8052}\\[3mm]
$^{2}$ Kitasato University,\,\, $^{3}$ Keio University,\,\, $^{4}$ Tokyo University\\[10mm]
\end{center}

\hrule
\vspace{3mm}
\noindent
\textbf{\large Abstract}:  A minor improvement is made to the calculation of the inhomogeneity term. The new calculation gives  better agreement with the observations by Daoud et al. and Cheng-Graessley-Melnichenko.
\vspace{0mm}
\begin{flushleft}
\normalsize{}\\[3mm]
\end{flushleft}
\hrule
\vspace{8mm}
\setlength{\baselineskip}{14pt}

\vspace{0mm}
\section{Introduction}
The basic equation for the excluded volume effects:
\begin{equation}
\frac{d G}{d\alpha}=\left(\mu_{c_{2\II}}-\mu_{c_{2I}}\right)\frac{d c_{2}}{d\alpha}+\frac{dG_{\text{elasticity}}}{d\alpha}=0\label{MIM-1}
\end{equation}
where $G$ is the Gibbs potential in the system and $\mu$ the chemical potential; the subscript $\II$ denotes a more concentrated region and $\I$ a more dilute region; $c_{2}$ denotes the concentration of solute molecules. $\mu$ is the quantity associated with the free energy of mixing\cite{Kazumi}.

The Gaussian approximation of an expanded polymer has the form: 
\begin{equation}
\Hat{p}(x,y,z)dxdydz=\left(\frac{\beta}{\pi\alpha^{2}}\right)^{3/2}\exp\left\{-\frac{\beta}{\alpha^{2}}\left[(x-a)^{2}+(y-b)^{2}+(z-c)^{2}\right]\right\}dxdydz\label{MIM-2}
\end{equation}
which, with $\delta V=dxdydz$, gives the segment concentration at the coordinates $(x, y, z)$:
\begin{align}
\Hat{C}(x, y, z)&=N\left(\frac{\beta}{\pi\alpha^{2}}\right)^{3/2}\sum_{\{a, b, c\}}\exp\left\{-\frac{\beta}{\alpha^{2}}\left[(x-a)^{2}+(y-b)^{2}+(z-c)^{2}\right]\right\}\notag\\
&=N\left(\frac{\beta}{\pi\alpha^{2}}\right)^{3/2}\HG(x, y, z)\label{MIM-3}
\end{align}
where the symbol $\Hat{}$ represents the approximate expression based on Eq. (\ref{MIM-2}). 

The first term in Eq. (\ref{MIM-1}) is obtained by using the free energy of mixing:
\begin{equation}
\mu_{c_{2}}=\left(\frac{\partial G_{mixing}}{\partial \Hat{C}}\right)_{T, P}\label{MIM-4}
\end{equation}
with
\begin{equation}
G_{mixing}=\,(kT/V_{1})\iiint (1-\Hat{v}_{2})\left\{\log\,\left(1-\Hat{v}_{2}\right)+\chi\Hat{v}_{2}\right\}dxdydz \label{MIM-5}
\end{equation}\\[-3mm]
where $V_1$ is the volume of a solvent molecule and $\Hat{v}_{2}=V_{2}\Hat{C}$ with $V_{2}$ being the volume of a segment\cite{Flory}. Hence, the difference of chemical potentials may be represented by a function of the difference of the volume fraction, $\Hat{v}$, of segments, namely,
\begin{equation}
\Hat{J}_\alpha^{\,k}=\iiint_{hill}\HG\hspace{0.3mm}^{\,k}\,dxdydz-\iiint_{valley}\HG\hspace{0.3mm}^{\,k}\,dxdydz \hspace{5mm} (k=1, 2, \cdots)\label{MIM-6}
\end{equation}\\[-2mm]
We make use of the lattice model and polymer molecules are arranged regularly on the sites of the simple cubic lattice having unit lengths, $p\times p\times p$. In the preceding works, we have defined $hill$ as an area enclosed by $[-p/4, p/4]$ for each axis and $valley$ as an area enclosed by $[p/4, 3p/4]$. However, the definition is not very accurate since the segment density is not isotropic. A more precise definition will be
\begin{equation}
\Hat{J}_\alpha^{\,k}=\iiint_{-p/4}^{p/4}\HG\hspace{0.3mm}^{\,k}\,dxdydz-\frac{1}{7}\left(\iiint_{-p/2}^{p/2}\HG\hspace{0.3mm}^{\,k}\,dxdydz-\iiint_{-p/4}^{p/4}\HG\hspace{0.3mm}^{\,k}\,dxdydz\right)\label{MIM-7}
\end{equation}

\begin{figure}[H]
\begin{center}
\includegraphics[width=15cm]{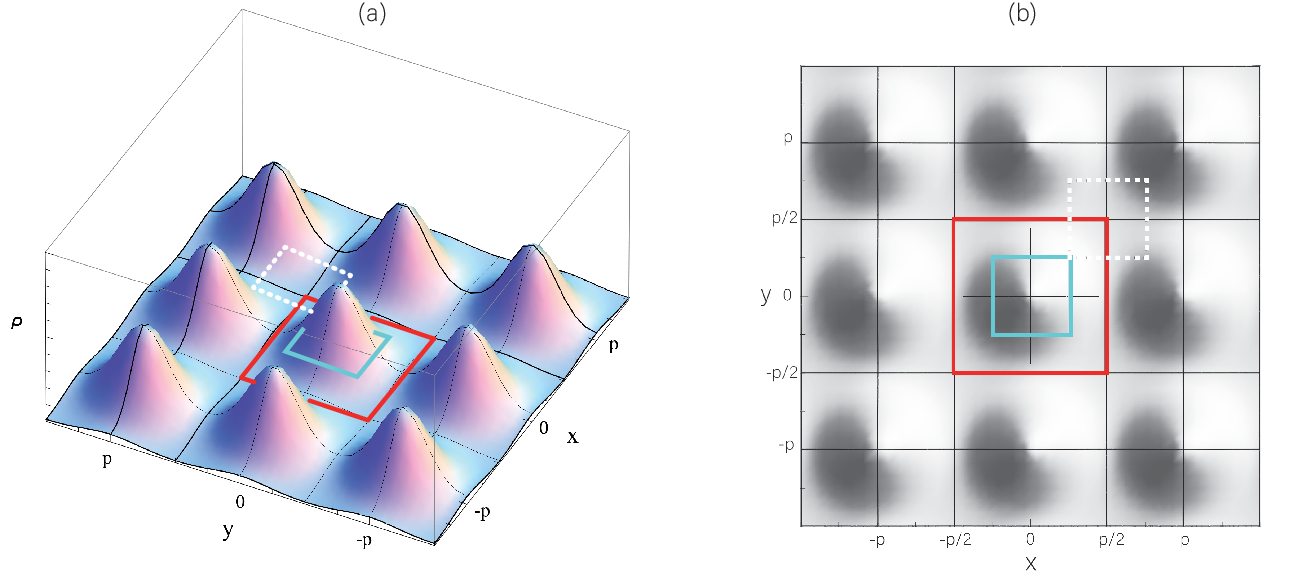}
\vspace{-2mm}
\caption{\setlength{\baselineskip}{11pt}An illustration of polymers arranged, in the equal interval $p$, at the sites of a plane $(x, y, 0)$ on the simple cubic lattice.}\label{Nenkai-1}
\end{center}
\vspace*{-6mm}
\end{figure}
\noindent Fig. \ref{Nenkai-1} illustrates polymer molecules on the simple cubic lattice and represents the segment density map on the coordinates $(x, y, 0)$, so that the peak height reflects the segment concentration [Fig. \ref{Nenkai-1}-(b) is a top view from the $z$-axis]. Each polymer is arranged in the center of a unit cell with the length $p\times p\times p$. The region shown by the blue square represents $hill$, while the region shown by the white dotted square represents the $valley$ as defined previously [see Eq. (\ref{MIM-11})]. In place of the previous definition, in the present calculation, the valley integral is defined by the equation $\frac{1}{7}\left(\iiint_{-p/2}^{p/2}\HG\hspace{0.3mm}^{\,k}\,dxdydz-\iiint_{-p/4}^{p/4}\HG\hspace{0.3mm}^{\,k}\,dxdydz\right)$, which will be an improvement since the depth of the valley is averaged out in conformity with the average circumstance of real polymer solutions.

Substituting Eqs. (\ref{MIM-3}) and (\ref{MIM-5}) into Eq. (\ref{MIM-1}), we have the expression derived precedently:

\begin{equation}
\alpha-1/\alpha =-\frac{1}{3V_{1}}\frac{\partial}{\partial\alpha}\iiint\left\{\left(1/2-\chi\right)\Hat{\mathscr{J}}^{2}+\frac{1}{6}\Hat{\mathscr{J}}^{3}+\cdots\right\}dxdydz\label{MIM-8}
\end{equation}
where $\Hat{\mathscr{J}}^{k}=\Hat{v}_{2, hill}^{k}-\Hat{v}_{2, valley}^{k}$ $(k=2, 3, \dots)$ together with\\[-2mm]
\begin{equation}
\Hat{v}_{2}=V_{2}\Hat{C}=V_{2}N\left(\frac{\beta}{\pi\alpha^{2}}\right)^{3/2}\HG(x, y, z)\label{MIM-9}
\end{equation}
$V_{2}$ can be equated with the volume of of a repeating unit of a polymer (see ref.\cite{Kazumi}), and\\[-3mm]
\begin{equation}
\HG(x, y, z)=\sum_{\{a, b, c\}}\exp\left\{-\frac{\beta}{\alpha^{2}}\left[(x-a)^{2}+(y-b)^{2}+(z-c)^{2}\right]\right\}\label{MIM-10}
\end{equation}\\[-4mm]
as defined in Eq. (\ref{MIM-3}).

Eq. (\ref{MIM-8}) can be solved numerically with the help of Eq. (\ref{MIM-7}). We show in the following $\Bar{\phi}$ and $N$ dependence of $\alpha$, and compare with the results by the previous definition:\\[-3mm]
\begin{equation}
\Hat{J}_\alpha^{\,k}=\iiint_{-p/4}^{p/4}\HG\hspace{0.3mm}^{\,k}\,dxdydz-\iiint_{-3 p/4}^{3 p/4}\HG\hspace{0.3mm}^{\,k}\,dxdydz\label{MIM-11}
\end{equation}

\section{Numerical Simulation}
\vspace{-2mm}
\subsection{The $\Bar{\phi}$ and $N$ dependence of $\alpha$ for linear polymers in good solvents}
The physical soundness of Eq. (\ref{MIM-8}) has already been proven through comparison with the observed data by Daoud and coworkers\cite{Daoud} and Cheng-Graessley-Melnichenko\cite{Graessley}. In this paper, we examine the difference in simulation results between the definitions (\ref{MIM-7}) and (\ref{MIM-11}), in light of the observed data by the same authors\cite{Daoud, Graessley}.

\begin{figure}[H]
\begin{center}
\begin{minipage}[t]{0.46\textwidth}
\begin{center}
\includegraphics[width=8cm]{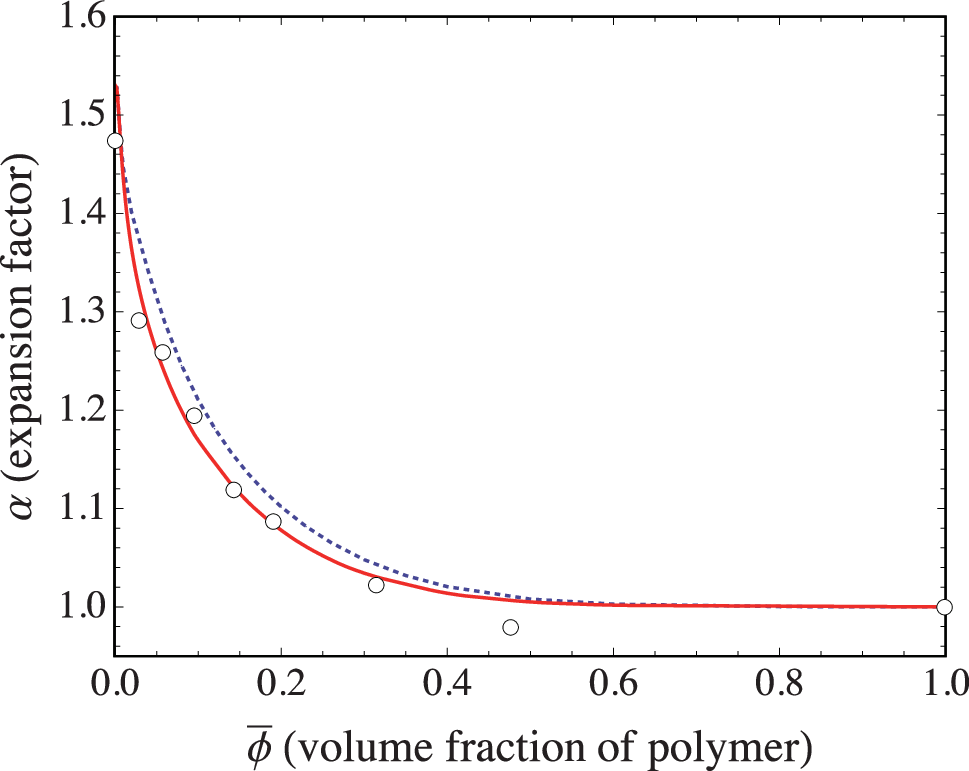}
\end{center}
\vspace{-2mm}
\caption{\setlength{\baselineskip}{11pt}The variations of the excluded volume effects as a function of $\Bar{\phi}$: PSt$-\text{CS}_{2}$ system ($N=1096$, $\chi=0.4$). Solid line: calculated by Eq. (\ref{MIM-7}); Dotted line: calculated by Eq. (\ref{MIM-11}). Open circles: observations by Daoud and coworkers; their observed value ($82\,\text{\AA}$) at $\Bar{\phi}=1$ is replaced by the mean value ($93\,\text{\AA}$) of all-workers'.}\label{Nenkai-2}
\end{minipage}
\hspace{10mm}
\begin{minipage}[t]{0.46\textwidth}
\begin{center}
\includegraphics[width=8cm]{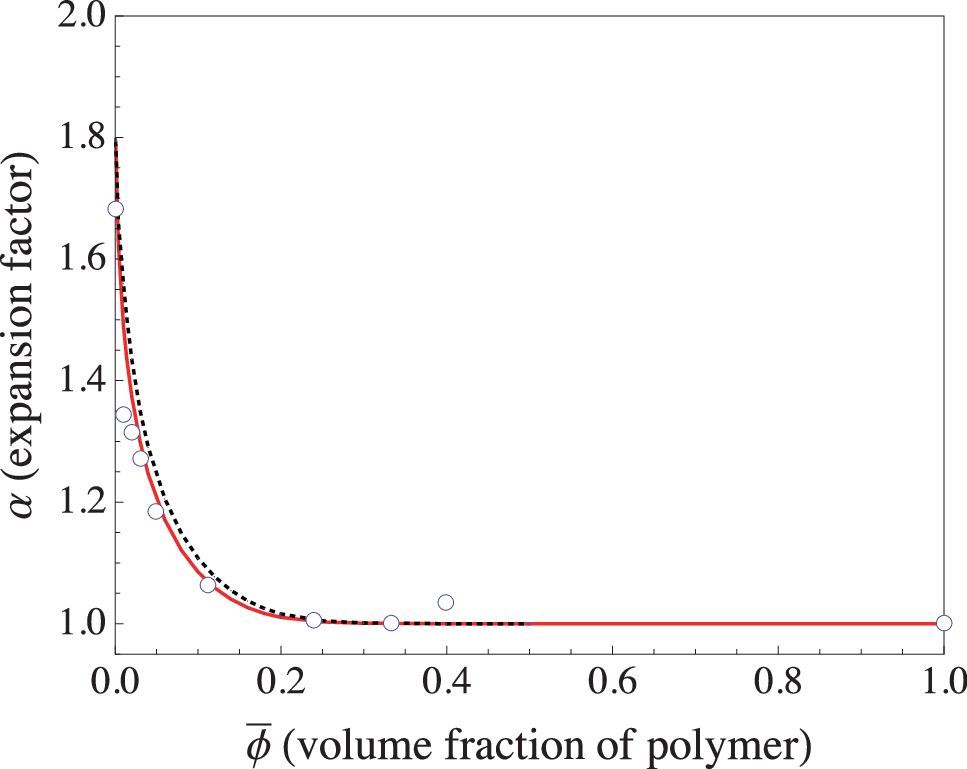}
\end{center}
\vspace{-2mm}
\caption{\setlength{\baselineskip}{11pt}The variations of the excluded volume effects as a function of $\Bar{\phi}$: PMMA$-\text{CHCl}_{3}$ system ($N=5900$, $\chi=0.3$). Solid line: results by Eq. (\ref{MIM-7}); Dotted line: results by Eq. (\ref{MIM-11}). Open circles: observations by Cheng-Graessley-Melnichenko.}\label{Nenkai-3}
\end{minipage}
\end{center}
\vspace*{0mm}
\end{figure}

In Fig. \ref{Nenkai-2}, $\Bar{\phi}$ (the mean volume fraction of polymers) dependence of $\alpha$ is plotted according to Eq. (\ref{MIM-8}) for the PSt$-\text{CS}_{2}$ system [$N=1096$, $V_1=100 \text{\AA}$ ($\text{CS}_{2}$), $V_2=165 \text{\AA}^3$, $\CF=10$, $\chi=0.4$, and $\Bar{l}=1.55 \text{\AA}$], while Fig. \ref{Nenkai-3} shows the corresponding $\alpha\,\, vs.\,\Bar{\phi}$ curves for the PMMA$-\text{CHCl}_{3}$ system [$N=5900$, $V_1=134 \text{\AA}$ ($\text{CHCl}_{3}$), $V_2=140 \text{\AA}^3$, $\CF=9.2$, $\chi=0.3$, and $\Bar{l}=1.56 \text{\AA}$]. For both cases, the solid lines are drawn by using Eq. (\ref{MIM-7}), and the dotted lines are by Eq. (\ref{MIM-11}). As one can see, the new definition (\ref{MIM-7}) yields smaller numerical values over the whole range of $\Bar{\phi}$ than the previous one (\ref{MIM-11}). These are reasonable results, since by averaging (deep and shallow) valley areas, the inhomogeneity term should decline to yield smaller $\alpha$'s. As mentioned in the previous study, on the other hand, $\alpha$ values in the two limits, the dilution limit ($\Bar{\phi}=0$) and the high concentration limit ($\Hat{J}_\alpha^{\,k}\rightarrow0$), are independent of the inhomogeneity term, so the alteration in the $valley$ integrals changes the radius of the curvature alone. 
\noindent Based on the above information, let us look at Figs. (\ref{Nenkai-2}) and (\ref{Nenkai-3}) again; it is seen that the new definition (\ref{MIM-7}) seems to lead to better agreement with the observed points.

Then, let us inspect how the two definitions make a difference for branched molecules.

\subsection{The $\Bar{\phi}$ and $N$ dependence of $\alpha$ for the $z=2$ polymer in good solvents}
For the present purpose, we use the $z=2$ polymer, the chemical structure of which is
\begin{figure}[H]
\begin{center}
\includegraphics[width=6cm]{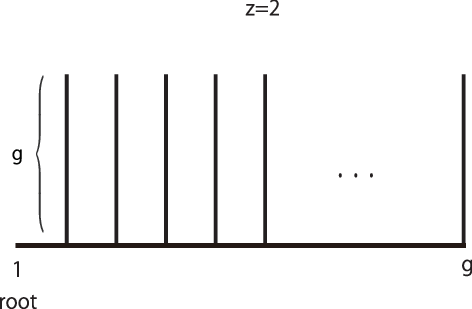}
\vspace{-2mm}
\caption{\setlength{\baselineskip}{11pt}The $z=2$ polymer.}\label{Nenkai-4}
\end{center}
\vspace*{-4mm}
\end{figure}
\noindent This polymer has the mean square of the radius of gyration:
\begin{equation}
\left\langle s_{N}^{2}\right\rangle_{0}=\frac{1}{6}\frac{(g^{2}-1)(4g-3)}{g^{2}}\,l^{2}\label{MIM-12}
\end{equation}
where $g$ is the generation number\cite{Kazumi}. Since the $z=2$ polymer obeys the relationship, $N=g^2$, we have $\nu_{0}=1/4$. 

\begin{figure}[H]
\begin{center}
\begin{minipage}[t]{0.46\textwidth}
\begin{center}
\includegraphics[width=7.5cm]{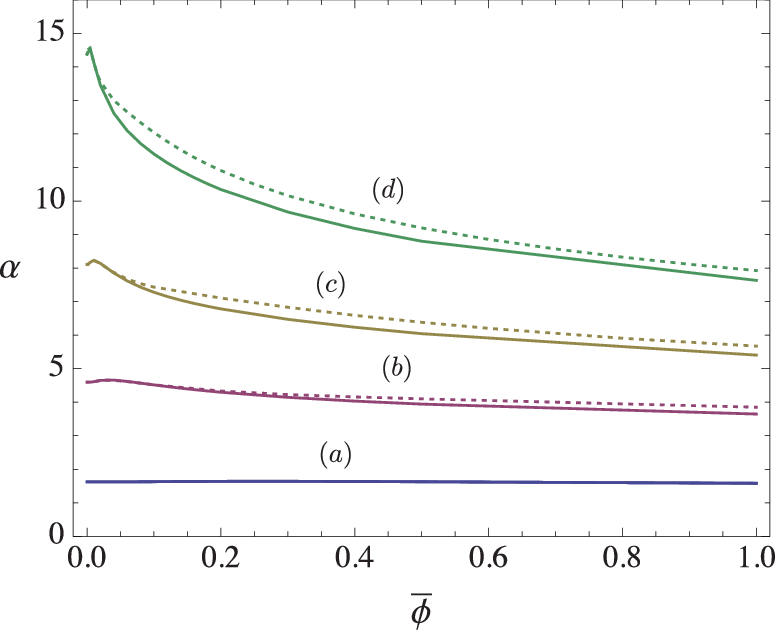}
\end{center}
\vspace{-2mm}
\caption{\setlength{\baselineskip}{11pt}The variations of the excluded volume effects as a function of $\Bar{\phi}$ for the $z=2-n$-\text{nonadecane} system ($\chi=0.2$):(a) $N=10^{2}$, (b) $N=10^{4}$, (c) $N=10^{5}$, (d) $N=10^{6}$. Solid line: results by Eq. (\ref{MIM-7}); Dotted line: results by Eq. (\ref{MIM-11}).}\label{Nenkai-5}
\end{minipage}
\hspace{10mm}
\begin{minipage}[t]{0.46\textwidth}
\begin{center}
\includegraphics[width=8.2cm]{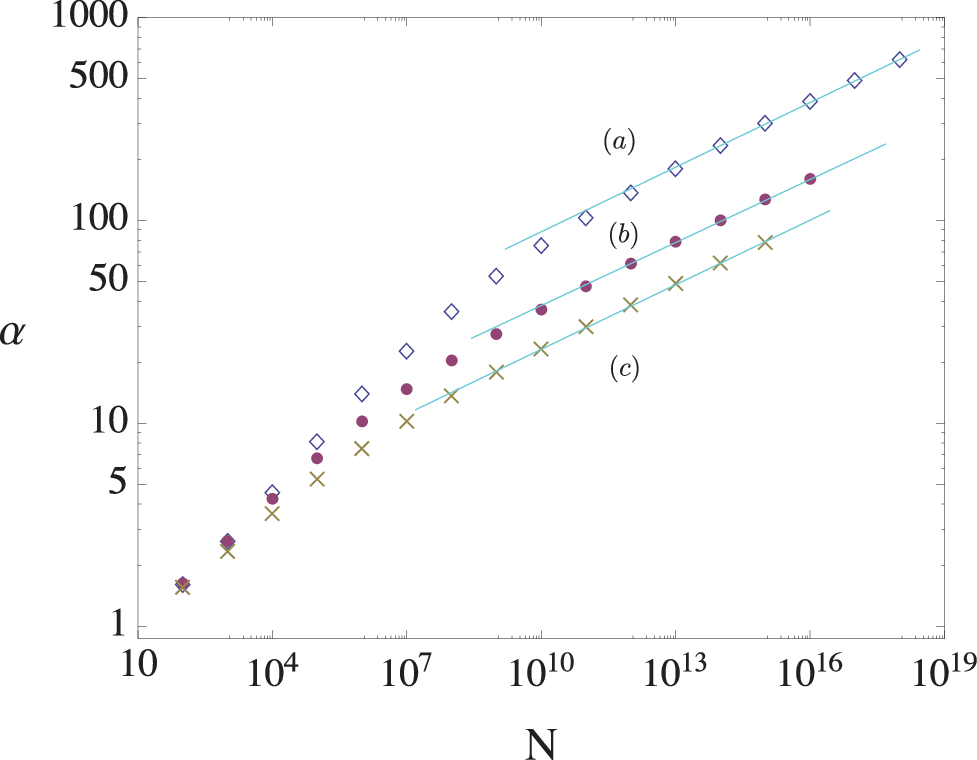}
\end{center}
\vspace{-2mm}
\caption{\setlength{\baselineskip}{11pt}The variations of the excluded volume effects as a function of $N$ for the $z=2-n$-\text{nonadecane} system ($\chi=0.2$): (a) $\Bar{\phi}=0.01$, (b) $\Bar{\phi}=0.2$, (c) $\Bar{\phi}=1$ (melt). Calculated by Eq. (\ref{MIM-7}).}\label{Nenkai-6}
\end{minipage}
\end{center}
\vspace*{0mm}
\end{figure}

Let a monomer unit of this polymer be composed of the methylene unit ($-\text{CH}_{2}-$: a segment). Then, the molecular mass is $M=14\,N+2$. We apply Eq. (\ref{MIM-8}) to this polymer. Employed parameters are $V_{1}=569\,\text{\AA}$ ($n$-nonadecane), $V_{2}=25\,\text{\AA}$, $\CF=7.7$, $\chi=0.2$, and $l=1.54\,\text{\AA}$. So, our system mimicks the PE$-n$-nonadecane system by Westermann, Willner, Richter, and Fetters\cite{Westermann}. Simulation results are displayed in Fig. \ref{Nenkai-5}. 

\hspace{3mm}
\noindent The essential features of the calculation results are the same as linear polymers in that they show lower values over the whole concentration range. 

\begin{figure}[H]
\begin{center}
\includegraphics[width=9cm]{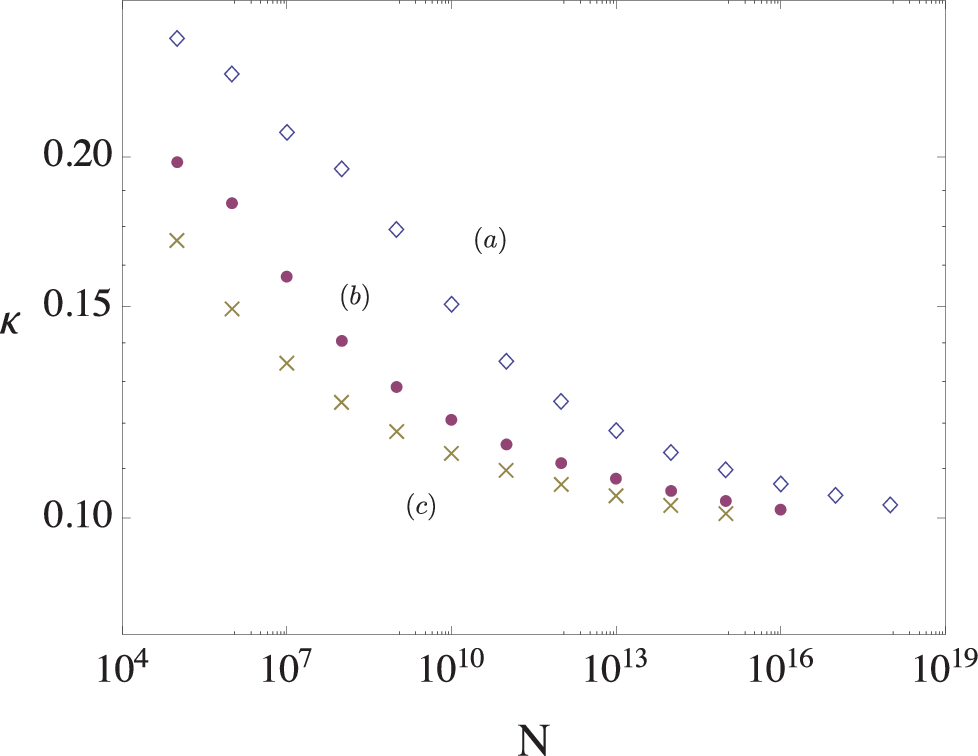}
\vspace{-2mm}
\caption{\setlength{\baselineskip}{11pt}$\kappa\,\, vs.\,N$ for the $z=2-n$-\text{nonadecane} system ($\chi=0.2$). The plot points were calculated according to Eq. (\ref{MIM-7}) for (a) $\bar{\phi}=0.01$, (b) $\bar{\phi}=0.2$, and (c) $\bar{\phi}=1$ (melt).}\label{Nenkai-7}
\end{center}
\vspace*{-4mm}
\end{figure}

A problem is whether or not the alteration of the valley integral from Eq. (\ref{MIM-11}) to Eq. (\ref{MIM-7}) makes an effect on the size exponent, $\nu$. To examine this, we have calculated the $N$ dependence of $\kappa$ in Figs. \ref{Nenkai-6} and \ref{Nenkai-7} for various concentrations, $\Bar{\phi} =0.01$, 0.2, and 1, where $\kappa$ is the exponent defined by $\alpha\propto N^\kappa$ so that $\kappa=\nu - \nu_0$.
As one can see, the observed gradients, $\kappa=\log\alpha /\log N$, approach $\doteq 0.1$ at $10^{15}- 10^{18}$ but are not still in the steady state for all $\Bar{\phi}$'s. It appears that the $\kappa$ values approach, in the event, 1/12, the critical packing density, as $N\rightarrow\infty$, the behavior of which is quite identical to the results with the previous definition (11). We expect, thus, the same $\nu\doteq 1/3$ for the whole concentration range of $0<\Bar{\phi}\le 1$.

\subsection{Conclusion}
The employment of the definition (\ref{MIM-7}) appears to lead to a minor improvement to the theory; i.e., it changes the radius of the curvature of the $\alpha\,\, vs.\,\Bar{\phi}$ lines, resulting in better agreement with the observations, but has no effect on the size exponent.

\newpage

\end{document}